\documentclass[aps,prb,superscriptaddress,amsmath,amssymb,twocolumn,amsfonts,floatfix, longbibliography]{revtex4-2}
\usepackage{graphicx}
\usepackage{epstopdf}
\usepackage[T1]{fontenc}
\usepackage[latin9]{inputenc}
\usepackage{amsbsy}
\usepackage{gensymb}
\usepackage{bm}

\usepackage[dvipsnames]{xcolor}
\definecolor{deepfuchsia}{rgb}{0.76, 0.33, 0.76}
\definecolor{electricpurple}{rgb}{0.75, 0.0, 1.0}
\usepackage[colorlinks=true,linktoc=page,linkcolor=blue,citecolor=magenta,urlcolor=electricpurple]{hyperref}
\usepackage{orcidlink}
\usepackage[normalem]{ulem}

\newcommand{\beq}{\begin{equation}}
\newcommand{\eeq}{\end{equation}}
\newcommand{\bea}{\begin{eqnarray}}
\newcommand{\eea}{\end{eqnarray}}
\newcommand{\non}{\nonumber}

\newcommand{\fig}[1]{Fig.~\ref{#1}}

\newcommand{\equref}[1]{Eq.~(\ref{#1})}

\newcommand{\secref}[1]{Sec.~\ref{#1}}
\newcommand{\figref}[1]{Fig.~\ref{#1}}

\newcommand{\appref}[1]{Appendix~\ref{#1}}

\renewcommand{\vec}[1]{\boldsymbol{#1}}

\begin{document}


\title{\textrm{Josephson diode effect in multichannel Rashba nanowires: Role of inter-subband coupling}}

\author{{Ardamon Sten}\,\orcidlink{0009-0006-2227-0695}}
\email[]{ardamons21@iitk.ac.in}
\affiliation{Department of Physics, Indian Institute of Technology, Kanpur 208016, India}

\author{{Sudeep Kumar Ghosh}\,\orcidlink{0000-0002-3646-0629}}
\email[]{skghosh@iitk.ac.in}
\affiliation{Department of Physics, Indian Institute of Technology, Kanpur 208016, India}

\date{\today}

\begin{abstract}
The Josephson diode effect (JDE) has attracted significant attention for enabling directional, dissipationless supercurrents, positioning Josephson junctions as promising building blocks for next-generation quantum devices. Hybrid semiconductor-superconductor nanowires provide an experimentally accessible platform for realizing the JDE and hosting Majorana bound states. However, most theoretical treatments assume the single-channel limit, whereas realistic nanowire devices are inherently multichannel due to transverse confinement. Here, we investigate the JDE in multichannel Rashba nanowire Josephson junctions, focusing on the role of inter-subband coupling. We show that subband hybridization qualitatively modifies both the topological phase diagram and the JDE response of the device. In contrast to the single-channel case, the topological phase is confined to a finite window of Zeeman fields, within which Majorana bound states strongly enhance the diode efficiency. Inter-subband coupling also enables a finite JDE even when the Zeeman field is aligned along the spin-orbit direction-- a mechanism absent in independent-channel and strictly one-dimensional nanowire systems. Furthermore, inter-subband coupling enhances spectral asymmetry and significantly increases the diode efficiency compared to single-channel junctions. These results identify inter-subband hybridization as a key ingredient for realizing and optimizing nonreciprocal superconducting transport in experimentally relevant hybrid nanowire Josephson junctions.

\end{abstract}

\maketitle

\section{Introduction}
Nonreciprocal superconducting transport has emerged as a central theme in modern quantum materials, offering a dissipationless route toward directional electronic functionalities. A primary manifestation of this behavior is the superconducting diode effect~\cite{He_2022,nadeem2023superconducting,Daido2022}, where the critical current differs for opposite bias directions. Microscopically, such nonreciprocity originates from the interplay of broken time-reversal and inversion symmetries, which can give rise to magnetochiral anisotropy and finite-momentum pairing states \cite{Noah2022,Legg2022,bhowmik2025optimizing,bhowmik2025field,samanta2025field,pal2022josephson}. An alternative route to intrinsic nonreciprocal transport can also be achieved in noncentrosymmetric superconductors via spontaneous time-reversal symmetry breaking~\cite{Ghosh2020a,Shang2022Weyl,Shang2018,Shang2020,Sajilesh2025time,kataria2026}. 

Josephson junctions (JJs) \cite{JOSEPHSON1962251, Likharev1979, Golubov2004} provide a particularly versatile and controllable platform to realize and probe this nonreciprocity. In this context, the superconducting diode effect manifests as the Josephson diode effect (JDE), characterized by an asymmetric current-phase relation (CPR), $I(\phi) \neq -I(-\phi)$, and hence unequal positive and negative critical currents. Such asymmetry typically arises when spin-orbit coupling (SOC) and an external magnetic field jointly break the relevant symmetries \cite{Davydova2022, Zhang2022}, inducing a phase-dependent asymmetry in the spectrum of the Andreev bound states (ABS).

The JDE has been investigated across a diverse range of platforms, including finite-momentum superconductors \cite{Davydova2022, Noah2022, Fu2024}, quantum-dot junctions \cite{Cheng2023, Debnath2024, Debnath_2025}, topological insulators \cite{Karabassov_2022,Lu2023, Wang2024, Fracassi2024, Huang2024}, altermagnets \cite{Cheng2024, Boruah2025, Jiang_2025,sharma_2025} and systems with SOC in the presence of a Zeeman field \cite{Reynoso2012, Liu2024, Meyer2024, Fu2024}. Furthermore, the JDE has been proposed in various geometries, such as kinked junctions \cite{Kopasov2021, Maiellaro2024} and SQUIDs \cite{Souto2022, Fominov2022, Ciaccia2023, Legg2023, Cuozzo2024, Qi2025}. Among these, semiconductor-superconductor hybrid nanowires are a promising platform as they naturally combine strong SOC, proximity-induced superconductivity, and magnetic tunability, while also hosting Majorana bound states (MBS) \cite{Oreg_2010,lutchyn_2010, Sau_2010, alicea_2010,Cayao2018} that can strongly influence the diode response. Existing theoretical works \cite{Cayao2024, Liu2024, Mondal2025} have shown that MBS can enhance the JDE, but they typically rely on an effective single-channel one-dimensional (1D) description. In realistic nanowires, however, transverse confinement leads to multiple occupied subbands and finite inter-subband coupling~\cite{Lutchyn2011,Antipov2018}. This multichannel structure is not merely a quantitative correction but can qualitatively modify both the topological properties and the transport response~\cite{Antipov2018,Carlos2025}. This raises a central question: how does inter-subband coupling in multichannel nanowire JJs reshape the topological phase diagram and the Josephson diode response, and can it enable qualitatively new mechanisms for nonreciprocal transport that are absent in single-channel systems?

In this work, we address this question by investigating the JDE in multichannel Rashba nanowire Josephson junctions, explicitly incorporating the role of inter-subband coupling. We show that subband hybridization qualitatively modifies the topological phase diagram and the nonreciprocal supercurrent, leading to a finite topological regime as a function of Zeeman field and a pronounced enhancement of the diode response. Most importantly, we find that inter-subband coupling enables a finite diode effect even when the Zeeman field is oriented purely along the spin-orbit direction, a feature absent in independent-channel and strictly one-dimensional limits. We further demonstrate that multichannel effects significantly enhance the diode efficiency compared to single-channel junctions. These results highlight the crucial role of multichannel hybridization in realistic devices and provide a pathway for optimizing nonreciprocal superconducting transport in hybrid nanowires.

The rest of the article is organized as follows: \secref{sec:model} introduces the model Hamiltonian for a cylindrical quasi-1D Rashba nanowire Josephson junction and details the topological properties of the proximitized system under an external magnetic field. In \secref{sec:Int_case}, we investigate the phase-dependent low-energy spectrum and non-reciprocal current-phase relationship (CPR) for subbands coupled via transverse SOC across various in-plane magnetic field orientations. The corresponding analysis for the case of independent subbands is presented in \secref{sec:ind_case} for comparison. The emergence and efficiency of the JDE are discussed in \secref{sec:efficiency}, followed by an examination of finite-temperature effects on the Josephson current and diode performance in \secref{sec:finite_temp}. We also extend the investigation for the case of multiple occupied subbands in \secref{sec:3_subbands}. In \secref{sec:rect_case}, we verify the robustness of our findings using a quasi-1D Rashba nanowire with hardwall confinement, before concluding with a summary and outlook in \secref{sec:conclusion}.

\section{Cylindrical Rashba nanowire Josephson junction model}
\label{sec:model}
\begin{figure}
    \centering
    \includegraphics[width=\linewidth]{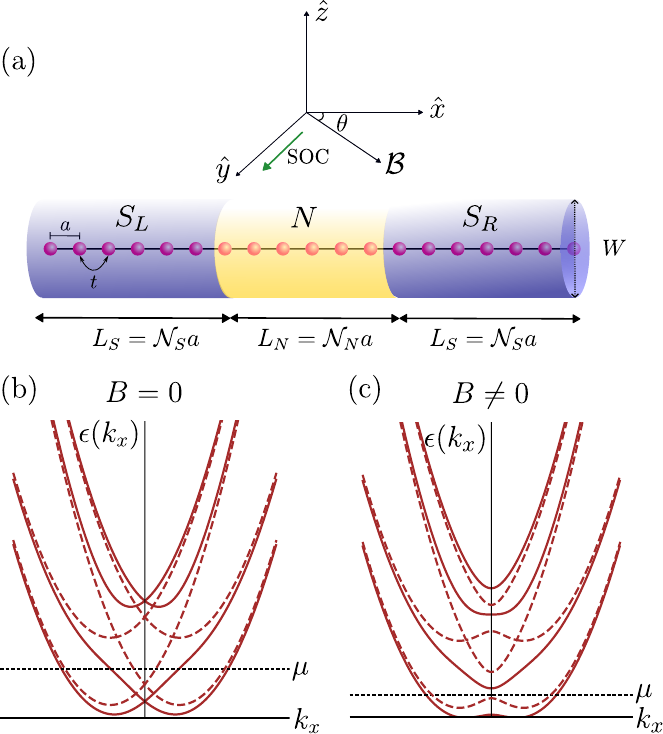}
    \caption{\textbf{Schematic of a multichannel Rashba nanowire Josephson junction:} (a) Device geometry showing a proximitized cylindrical Rashba nanowire of diameter $W$ with spin-orbit coupling along the $y$-direction. $S_{L/R}$ and $N$ denote the proximitized superconducting and normal regions, subject to a uniform external Zeeman field $\vec{\mathcal{B}}$ applied at an angle $\theta$ with the $x$-axis throughout the junction. The junction is discretized into an effective 1D lattice with spacing $a$ and hopping energy $t$; the superconducting and normal regions consist of $\mathcal{N}_S$ and $\mathcal{N}_N$ lattice points, yielding lengths $L_S = a \mathcal{N}_S$ and $L_N = a \mathcal{N}_N$. (b, c) Normal-state dispersion of the quasi-1D nanowire illustrating spin-split states for the two lowest subbands. Solid and dashed lines represent interacting and non-interacting subbands, respectively. At $B=0$, the dispersion remains spin-degenerate at $k_x=0$, whereas a finite field ($B \neq 0$) opens a Zeeman gap, entering the helical regime where spin-momentum locking ensures opposite spins at the two Fermi points. Here, $\mu$ denotes the chemical potential.
   }
    \label{fig:schematic}
\end{figure}

We consider a quasi-1D cylindrical Rashba nanowire oriented along the $x$-axis, with harmonic confinement in the transverse ($y$- and $z$-) directions. The Josephson junction is realized in a semiconductor-superconductor heterostructure, where the Rashba nanowire is proximitized by two $s$-wave superconductors separated by a distance $L_N$. The segments of the nanowire in contact with the superconductors acquire superconductivity via the proximity effect, while the intermediate region remains normal, thereby forming an effective superconductor-normal-superconductor (SNS) junction, as schematically illustrated in \fig{fig:schematic}(a). A uniform magnetic field $\vec{\mathcal{B}}$ is applied across the entire junction at an angle $\theta$ with respect to the $x$-axis.

Proximity-induced superconductivity in the Rashba nanowire can be described by the low-energy Bogoliubov-de Gennes (BdG) model Hamiltonian~\cite{bhowmik2025optimizing}
\begin{align}
    H_{\text{BdG}} &= (H_0-\mu)\tau_z + H_R\tau_z + H_Z + H_S,
    \label{eq:nanowire_hamil}
\end{align}
where $\mu$ is the chemical potential of the system. $H_0$ is the single-particle Hamiltonian of the quasi-1D nanowire:
\begin{equation}
    H_0 = \frac{p_x^2 + p_y^2 + p_z^2}{2m^*} + U_c(y,z).
\end{equation}
Here, $m^*$ is the electron's effective mass, $p_{\text{i}}=i\hbar\partial_{\text{i}}$ is the momentum operator along the corresponding directions and $U_c(y,z)=m^* \omega^2_0(y^2+z^2)/2$ is the harmonic confinement potential along the transverse axes with $\omega_0$ being the angular frequency of the harmonic potential. The angular frequency is related to the diameter of the nanowire as $W=2\sqrt{\hbar/(m^*\omega_0)}$ \cite{Park2017}. The confinement gives rise to subbands with discrete energy levels $\varepsilon_{n_y,n_z}=\hbar \omega_0(n_y+n_z+1)$ where $n_y,n_z=0,1,2\ldots$ are quantum numbers. $H_R$ and $H_Z$ are the Rashba SOC and Zeeman terms respectively and are given by
\begin{align}
    H_R&=-\alpha p_x\sigma_y + \alpha p_y \sigma_x,\\
    H_Z &= B_x \sigma_x \tau_z+ B_y \sigma_y \tau_0,\nonumber
\end{align}
where $\alpha$ is the SOC strength. $B_x=B \cos{\theta}$ and $B_y=B \sin{\theta}$ are the $x$ and $y$ components of the Zeeman field $B=g\mu_B|\mathcal{B}|/2$ due to the applied magnetic field $\mathcal{B}$ in the $x-y$ plane at an angle $\theta$ with the $x$-axis. Here, $g$ is the Lande g-factor and $\mu_B$ is the Bohr magneton. $\sigma_i$ and $\tau_i$ are the Pauli matrices in spin space and Nambu space respectively. The induced-superconductivity in the nanowire is described by
\begin{equation}
    H_S=\Delta [\sin{(\phi)}\sigma_y\tau_x + \cos{(\phi)}\sigma_y\tau_y ],
\end{equation}
where we have assumed strong proximity effect such that the induced gap $\Delta$ is equal to that of the parent s-wave superconductor and $\phi$ is the coherent phase of the superconducting state of the nanowire.

Tuning the chemical potential so that only the two lowest subbands are relevant at low energies, we project \equref{eq:nanowire_hamil} onto this subspace to obtain the effective Hamiltonian~\cite{Lutchyn2011,Park2017}:
\begin{align}
    \mathcal{H}_{\text{BdG}}^{\text{eff}}&=\frac{1}{2}\int dx \Psi^{\dagger}(x)H^{\text{eff}}_{\text{BdG}} \Psi(x), \non\\
    H^{\text{eff}}_{\text{BdG}}&=(H'_0-\mu)\tau_z + H'_R\tau_z + H'_Z + H_S.
    \label{eq:effective_hamil}
\end{align}
Here, the Hamiltonian is written in the basis $\Psi(x)=(\psi^e(x), \psi^h(x))^T$ where $\psi^e(x)=(\psi_{0\uparrow},\psi_{0\downarrow},\psi_{1\uparrow},\psi_{1\downarrow})^T$ and $\psi^h(x)=(\psi^{\dagger}_{0\downarrow},\psi^{\dagger}_{0\uparrow},\psi^{\dagger}_{1\downarrow},\psi^{\dagger}_{1\uparrow})^T$. The index $\text{j}=0,1$ on $\psi^{e/h}_{\text{j,s}}$ labels the two relevant subbands and $\text{s}=\uparrow, \downarrow$ denote the spins. The different terms in the projected Hamiltonian are given as
\begin{align}
    H'_0&=\frac{p^2_x}{2m^*} + \frac{\
    \varepsilon_-}{2}(1-\rho_z) \non\\
    H_R'&=-\alpha p_x \tilde{\sigma}_z + \delta_c \tilde{\sigma}_y\rho_y ,\\
    H_Z'&=B_x\tilde{\sigma}_y \tau_0 + B_y\tilde{\sigma}_z \tau_z,\non
\end{align}
where $\varepsilon_-=\varepsilon_{10}-\varepsilon_{00}$ is the subbands energy difference, $\delta_c=\sqrt{2}\hbar \alpha/W$ is the inter-subband coupling via SOC. The Pauli matrices $\tilde{\sigma}_{x,y,z}$ act on the projected spin space and ${\rho}_{x,y,z}$ act on the space of the two subbands.

\begin{figure}
    \centering
    \includegraphics[width= \linewidth]{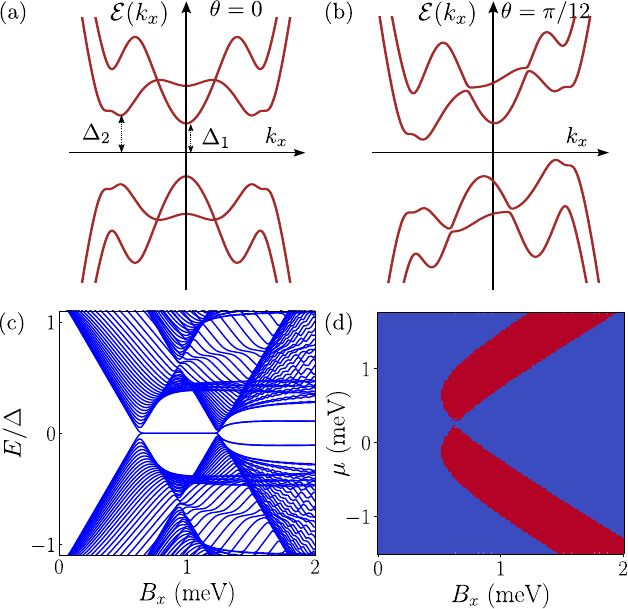}
    \caption{\textbf{Low-energy Bogoliubov quasiparticle spectrum and topological phase diagram of the cylindrical superconducting Rashba nanowire: } Evolution of the Bogoliubov quasiparticle spectrum of the nanowire as a function of momentum $k_x$ under an external field $B=1$ meV for distinct orientations of the Zeeman field: (a) $\theta=0$ and (b) $\theta=\pi/12$. (c) The low-energy spectrum as a function of the Zeeman field ($B_x$) applied along the $x$ direction at a fixed $\mu=-0.5$ meV, the zero-energy MBS emerging at one-subband occupancy hybridize to form trivial non-zero energy states at two-subband occupancy and (d) the corresponding topological phase diagram as a function of the chemical potential $\mu$ and Zeeman field $B_x$ where the blue and red regions represent the topologically trivial and non-trivial phases respectively.}
    \label{fig: dispersion_topology}
\end{figure}
Before we proceed to study the SNS multichannel junction, we first investigate the effects of the external Zeeman field and its different orientations on the low-energy spectrum of the proximitised nanowire with a finite interband coupling $\delta_c$. The normal-state energy dispersion in momentum space in the absence of a magnetic field features spin-split states of the two lowest subbands with a spin degeneracy at $k_x=0$ (\figref{fig:schematic}(b)), a finite magnetic field lifts the spin degeneracy and opens a Zeeman gap at $k_x=0$ (\figref{fig:schematic}(c)). When the chemical potential lies within this gap, there are only two Fermi points which have opposite spin and hence opposite helicity due to spin-momentum locking, thus this energy window is called the `helical regime'. For the real-space simulations, we consider realistic parameters \cite{lutchyn2018majorana} for our calculations: $m^*=0.015m_e$, $\Delta=0.5$ meV, $\hbar\alpha=40$ meV nm and $W=200$ nm. We present the dispersion of the BdG Hamiltonian in \equref{eq:effective_hamil} in momentum space in \figref{fig: dispersion_topology}. When the Zeeman field is applied along the transport direction ($\theta=0$), the spectrum is symmetric about $k_x=0$ and two induced gaps $\Delta_1$ and $\Delta_2$ open up at low and high momenta, respectively, as shown in \figref{fig: dispersion_topology}(a). In \figref{fig: dispersion_topology}(b), the Zeeman field is oriented at an angle $\theta=\pi/12$ and we notice that the presence of a non-zero $B_y$ component tilts the spectrum and renders it asymmetric about $k_x=0$, this asymmetry is crucial for observing diode effect as we shall see later. Another effect of a non-zero $\theta$ is the reduction of the high-momentum gap $\Delta_2$ while maintaining constant $\Delta_1$ as $\theta$ increases, and at a certain critical angle $\theta_c$ the gap vanishes and becomes negative for higher values of $\theta$. These behaviors are qualitatively similar to the case of a pure 1D nanowire junction \cite{Mondal2025}. To understand the topology of the system, we apply the magnetic field along the $x$ direction only since $B_x$ drives the topology of the system and present the low-energy spectrum as a function of the Zeeman energy, for a fixed chemical potential of $\mu=-0.5$ meV such that only one subband is occupied in the helical regime, in \figref{fig: dispersion_topology}(c). At lower values of the Zeeman field, the spectrum is fully gapped and as the Zeeman field increases the spectrum gap reduces and reaches the first critical point where a gap closing occurs which signifies a topological phase transition from a trivial to non-trivial topological phase. Beyond the first critical point, a pair of zero-energy states (ZES) emerge which correspond to a pair of MBS located at the ends of the nanowire. Furthermore, as the Zeeman field is further increased, a second critical point is reached where the gap reopens and the system undergoes a second phase transition from the topological to the trivial phase where the two MBS emerge as finite energy modes. This is a direct consequence of the spectral even-odd effect \cite{San-Jose2014, Stanescu_2011, Potter_2011} where the MBS are observed only at odd number of occupied subbands, beyond the second critical point the number of occupied subbands changes from one to two, therefore the two MBS hybridize to form two trivial non-zero-energy states. This signifies that a multichannel superconducting nanowire supports only a finite topological regime due to hybridization of the spectrum of the two subbands, this characteristic is unique to the multichannel case as compared to the single-channel where there is only one phase transition from trivial to topological at the critical field $B_c=\sqrt{\mu^2+\Delta^2}$. We also note that the critical field for topological phase transition is no longer given by $\sqrt{\mu^2+\Delta^2}$ in the case of multichannels with finite inter-subband coupling. To further demonstrate the finite topological regime of the multichannel nanowire, we present the topological phase diagram in \figref{fig: dispersion_topology}(d) as a function of the Zeeman energy and chemical potential where the blue and red regions represent the trivial and topological regimes respectively. The topological regime strongly depends on the chemical potential and Zeeman field; the two branches of the topological regime at low and high chemical potential correspond to the non-trivial phase arising from the occupancy of one and three subbands, respectively. In both the topological branches, further increasing the Zeeman field results in a transition back to the trivial regime due to change of occupancy from odd to even number of subbands, thus featuring a finite topological regime. This is in stark contrast to the single-channel case where the topological phase exists in the region bounded by $B\ge\sqrt{\mu^2 + \Delta^2}$.

We now proceed to investigate the JDE on the SNS multichannel nanowire junction. We discretize the Hamiltonian in \equref{eq:effective_hamil} into a tight-binding 1D lattice with an effective lattice constant $a=10$ nm. The left and right superconducting regions $S_{L/R}$ are characterized by their pair potentials $\Delta~e^{\mp i\phi/2}$ with $\mp \phi/2$ being their corresponding superconducting phases respectively, such that the phase difference across the junction is $\phi$. The lattice Hamiltonian of the junction is then given by
\begin{align} \label{eq:SNS_hamil}
    H_{\text{SNS}}&= \sum_ic^{\dagger}_{i,\rho}h_{ii}c_{i,\rho} + \sum_{\langle ij\rangle} c^{\dagger}_{i,\rho}T_{ij}c_{i,\rho} \non\\ 
    &+ \sum_{i} \Delta_{\alpha} e^{i\phi_\alpha}(c^{\dagger}_{i,\rho,\uparrow}c^{\dagger}_{i,\tau,\downarrow}-c^{\dagger}_{i,\rho,\downarrow}c^{\dagger}_{i,\rho,\uparrow}) + \textrm{h.c},
\end{align}
where $i$ labels the site index, $\rho=0,1$ labels the two subbands, $c^{\dagger}_{i\sigma}$ and $ c_{i\sigma}$ are the creation and annihilation operator of an electronic state at site $i$ with spin $\sigma={\uparrow,\downarrow}$. $\alpha=S_L,N,S_R$ labels the left superconductor, normal region and right superconductor respectively such that the pairing potentials $\Delta_{S_L,S_R}=\Delta$ in the two superconducting regions, $\Delta_{N}=0$ for the normal region and the corresponding phases $\phi_{S_L,S_R}$ take the forms as given above. The onsite ($h_{ii}$) and hopping ($T_{ij}$) terms are respectively given by
\begin{align}
    h_{ii}&=(2t-\mu) + E_-(1-\rho_z)+\delta_c \sigma_y \rho_y + B_x\sigma_y + B_y\sigma_z, \non\\
    T_{ij}&= -t\sigma_0 + it_{\textrm{SO}}\sigma_z,
\end{align}
where $t=\hbar^2/(2m^*a^2)$ is the hopping energy, $t_{\textrm{SO}}= \hbar\alpha/(2a)$ is the Rashba SOC hopping energy, and $\mu$ is the chemical potential which is kept the same for all regions. We investigate the JDE in the two distinct cases of: (i) interacting channels and (ii) independent channels in the next sections.

\section{Interacting channel case}
\label{sec:Int_case}
We first consider the case where the subbands are coupled via SOC and thus $\delta_c\ne0$. We consider a short and transparent junction throughout this article (unless explicitly mentioned otherwise) where the length of the normal region is $L_N=20$ nm and the superconductor length is taken to be $L_S=2~\mathrm{\mu m}$ (for junctions with longer normal region refer to \appref{app:long_junction}). Next, we numerically diagonalise the Hamiltonian in \equref{eq:SNS_hamil} according to the mentioned parameters and present the corresponding low-energy spectrum and current-phase relationship (CPR) in  the following subsections. 

\subsection{Low-energy spectrum}
\label{Int_spectrum}
\begin{figure}
    \centering
    \includegraphics[width=\linewidth]{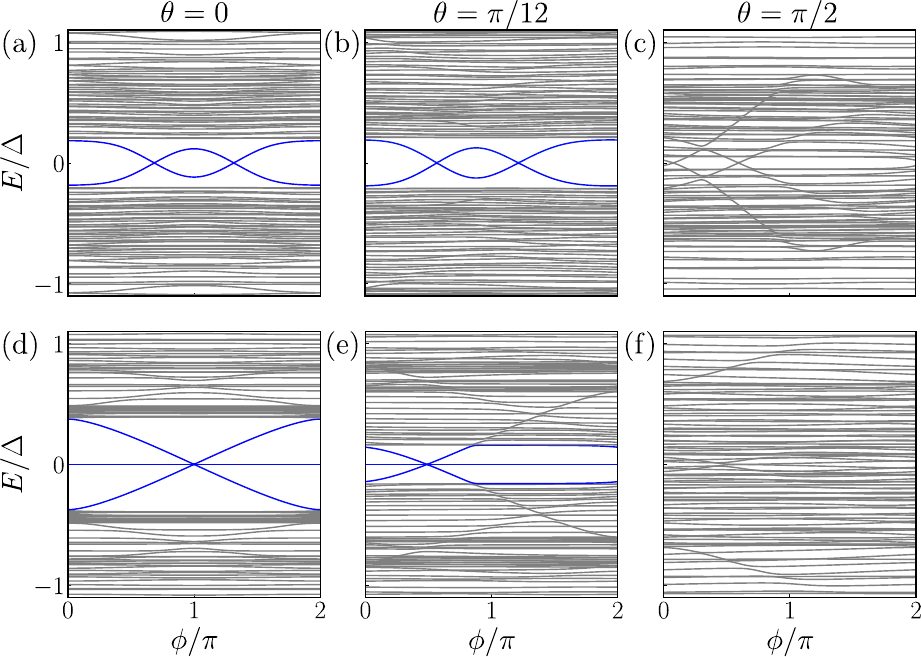}
    \caption{\textbf{Andreev spectrum of the interacting multichannel nanowire SNS junction with a uniform Zeeman field:}  The low-energy spectrum as a function of the phase difference $\phi$ for different orientations of the Zeeman field for: (a-c) the topologically trivial phase (B=0.5 meV) with $\mu=0.2$ meV and (d-f) the topologically non-trivial phase (B=1 meV) with $\mu=-0.5$ meV. The length of the superconductor and normal regions are taken to be $L_S=2~\mathrm{\mu m}$ and $L_N=20$ nm respectively.}
    \label{fig:SNS_spectrum_int}
\end{figure}
The low-energy spectrum is presented in \figref{fig:SNS_spectrum_int} for two values of the Zeeman field: (i) $B=0.5$ meV and (ii) $B=1$ meV corresponding to the trivial and topological phases respectively. In Figs. \ref{fig:SNS_spectrum_int}(a-c), we plot the spectrum in the trivial phase for three distinct values of $\theta$. For $\theta=0$, the spectrum is completely asymmetric about $\phi=\pi$ since $B_y=0$, and within the induced gap there are ABS with the lowest two energy states (blue curves) forming trivial zero-energy crossings on either side of $\phi=\pi$. Additionally, quasi-continuum states exist between the induced gap $\Delta_1$ and the parent gap $\Delta$ as shown in \figref{fig:SNS_spectrum_int}(a). As the angle is varied to $\theta\ne 0$, the $B_y$ component acquires a non-zero value, and as a result of which, a weak asymmetry is introduced in the spectrum as seen for $\theta=\pi/12$ in \figref{fig:SNS_spectrum_int}(b). As $\theta$ is increased further the spectrum changes drastically and at $\theta=\pi/2$ (\figref{fig:SNS_spectrum_int}(c)), the ABS disappear due to the reduction of the high-momentum gap $\Delta_2$ which even becomes negative beyond the critical angle $\theta_c$ as mentioned in \secref{sec:model}, thus reducing the induced gap and introducing quasi-continuum levels in the ABS regime. Figs. \ref{fig:SNS_spectrum_int}(d-f) show the corresponding spectrum in the topological phase; at $\theta=0$ the four lowest ABS (blue curves in \figref{fig:SNS_spectrum_int}(d)) develop non-trivial zero energy states. The first two ABS form a pair of ZESs which are non-dispersive with $\phi$ and correspond to a pair of MBS located at the nanowire ends, the next two ABS develop a zero-energy crossing at $\phi=\pi$ corresponding to another pair of MBS which are located at the two interfaces of the junction. At $\theta=\pi/12$ the $B_y$ component creates a strong asymmetry in the spectrum and shifts the MBS away from $\phi=\pi$, moreover, the spectrum asymmetry is enhanced by the presence of the MBS in the topological phase as seen in \figref{fig:SNS_spectrum_int}(e). Increasing $\theta$ further results in the gap $\Delta_2$ becoming negative which introduces trivial quasi-continuum levels along with MBS thus the MBS are no longer topologically protected. Interestingly, we also notice that there is an asymmetry in the spectrum for $\theta=\pi/2$ as seen in Figs. \ref{fig:SNS_spectrum_int}(c) and (f) for the interacting subbands (for more discussion refer to \appref{app:transverse_field_asymmetry}). This implies that a non-reciprocal CPR exists even in the absence of the $B_x$ component as we will see in the following subsection.

\subsection{Current-phase relationship}
\begin{figure}
    \centering
    \includegraphics[width=\linewidth]{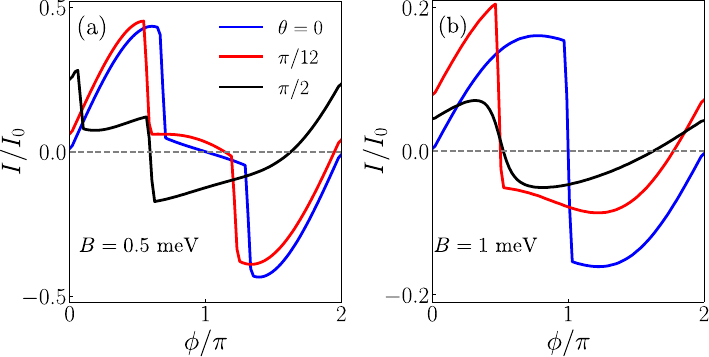}
    \caption{\textbf{Current-phase relationship of the interacting multichannel nanowire SNS junction with a uniform Zeeman field:}  The Josephson current at zero temperature as a function of the phase difference $\phi$ for different orientations of the Zeeman field in: (a) the topologically trivial phase (B=0.5 meV) with $\mu=0.2$ meV and (b) the topologically non-trivial phase (B=1 meV) with $\mu=-0.5$ meV. Here, $I_0=e\Delta/\hbar$ and the length of the superconductor and normal regions are taken to be $L_S=2~\mathrm{\mu m}$ and $L_N=20$ nm respectively.}
    \label{fig:SNS_CPR_int}
\end{figure}
The Josephson current at a temperature $T$ is obtained from the energy spectrum of the junction using the formula \cite{beenakker1992three},
\begin{equation}\label{current_formula}
   I(\phi)=-\frac{e}{\hbar}\sum_{E_n>0}\frac{dE_n(\phi)}{d\phi}\tanh{\left(\frac{E_n(\phi)}{2k_BT}\right)},
\end{equation}
where $E_n(\phi)$ denotes the phase-dependent energy levels of the Hamiltonian in \equref{eq:SNS_hamil}, $e$, $\hbar$ and $k_B$ are the electron's charge, reduced Planck's constant and the Boltzmann constant respectively. We present the CPR at zero temperature for different orientations of the Zeeman field in the trivial and topological phases, corresponding to the different energy spectra in subsection \ref{Int_spectrum} in \figref{fig:SNS_CPR_int}. \figref{fig:SNS_CPR_int}(a) shows the CPR in the trivial phase ($B=0.5$ meV) for the three distinct values of $\theta$. At $\theta=0$, the CPR is symmetric about $\phi=\pi$ and exhibits a sharp drop in the current magnitude without changing sign at the zero-crossings of the ABS. The absence of a sign change at the ABS crossings shows the finite contribution of the continuum levels in addition to the ABS. For $\theta\ne0$, the weak asymmetry in the spectrum due to presence of a finite $B_y$ component is reflected in the asymmetric CPR where the positive and negative critical currents ($I_c^{\pm}$), defined as $I_c^{\pm}=\text{max}[\pm I(\phi)]$, have different magnitudes and $I(\phi)\ne -I(-\phi)$ as shown by the red ($\theta=\pi/12$) and black ($\theta=\pi/2$) curves in \figref{fig:SNS_CPR_int}(a), signaling the presence of diode effect. \figref{fig:SNS_CPR_int}(b) shows the CPR for the same values of $\theta$ in the topological phase ($B=1$ meV). The CPR for $\theta=0$ (blue curve) is symmetric around $\phi=\pi$ and features a sign change and sharp discontinuity at $\phi=\pi$ where the two pairs of MBS are formed. The red curve in \figref{fig:SNS_CPR_int}(b) represents the CPR for $\theta=\pi/12$; the strong asymmetry in the spectrum results in an asymmetric CPR where the sharp discontinuity is shifted away from $\phi=\pi$ due to the shifting of the MBS as seen in \figref{fig:SNS_spectrum_int}(e). Moreover, the difference between the critical currents $I_c^+$ and $I_c^-$ is more pronounced in the topological regime. This means that the JDE is enhanced in the topological regime due to the presence of non-local MBS. When the Zeeman field lies wholly along the SOC direction ($\theta=\pi/2$), we notice that the CPR and critical currents asymmetry persists even in the absence of the $B_x$ component as shown by the black curve in \figref{fig:SNS_CPR_int}(b), signifying that the JDE can be realized with just the transverse field alone for interacting channels \cite{santra2026superconducting}.

\section{Independent channel case}
\label{sec:ind_case}
\begin{figure}
    \centering
    \includegraphics[width= \linewidth]{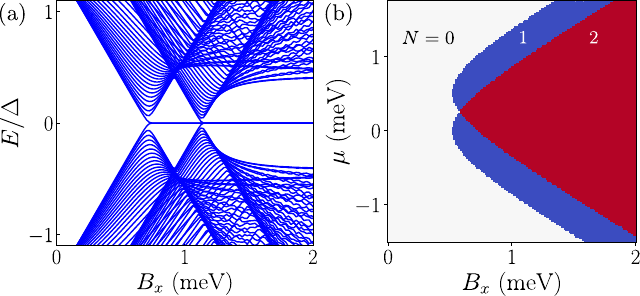}
    \caption{\textbf{Low-energy spectrum and topological phase diagram of the non-interacting multichannel superconducting Rashba nanowire:}
    (a) The low-energy spectrum as a function of the Zeeman field ($B_x$) applied along the $x$-direction and (b) the topological phase diagram shown in the ($\mu,B_x$) parameter space with the white, blue and red regions representing the number of MBS pairs ($N$) in the trivial and topological regimes respectively.
    }
    \label{fig:BdG_spectrum_ind}
\end{figure}

Next, we consider the case where the subbands coupling via SOC is negligible and thus $\delta_c=0$ and the subbands behave as independent channels. This is realised in systems with low SOC and nanowires with large diameter such that $\hbar\alpha\ll W$. To understand how the spectrum and the topology of the system modifies for the case of independent channels, we consider a superconductivity-proximitised multichannel nanowire in the presence of an external Zeeman field and present its low-energy spectrum and topological phase-diagram in \figref{fig:BdG_spectrum_ind}. The evolution of the low-energy spectrum with the external Zeeman field applied along the transport direction ($x$-axis) as shown in \figref{fig:BdG_spectrum_ind}(a) exhibits a stark contrast to that of the interacting-channel case (see \figref{fig: dispersion_topology}(c)). The spectrum of each subband is qualitatively similar to the pure 1D case and the subbands spectra coexist without hybridization with each other. The spectrum is initially gapped for lower values of the Zeeman field and later becomes gapless as each subband reaches its corresponding critical field where the topological phase transition from trivial to non-trivial occurs, the critical fields for the first and second subband are respectively given by $B_{c1}=\sqrt{\Delta^2 + \mu^2}$ and $B_{c2}=\sqrt{\Delta^2 + (\mu-E_-)^2}$ where $E_-$ is the subband energy difference. The coexistence of the topological regimes of the two subbands without any hybridization implies that by tuning the Zeeman field to a value $B\ge B_{c2}$, two pairs of MBS can be observed with each pair coming from each of the subbands. This is further demonstrated by the topological phase diagram in \figref{fig:BdG_spectrum_ind}(b) where the number of MBS is shown in the $(\mu,B_x)$ parameter space, the white region corresponds to the trivial phase where no MBS are observed while the blue region corresponds to the topological phase which hosts one pair of MBS and the red region corresponds to the overlap of the topological regimes of the two independent subbands resulting in a topological phase which hosts two pairs of MBS. With the understanding of the different topological regimes of the independent-channel limit, we now proceed to study the JDE arising in the SNS junction in these different topological regimes.
We keep the same parameters as in \secref{sec:Int_case} and proceed in a similar way and present the corresponding low-energy spectrum and current-phase relationship (CPR) in the following subsections. 

\subsection{Low-energy spectrum}

\begin{figure}
    \centering
    \includegraphics[width=\linewidth]{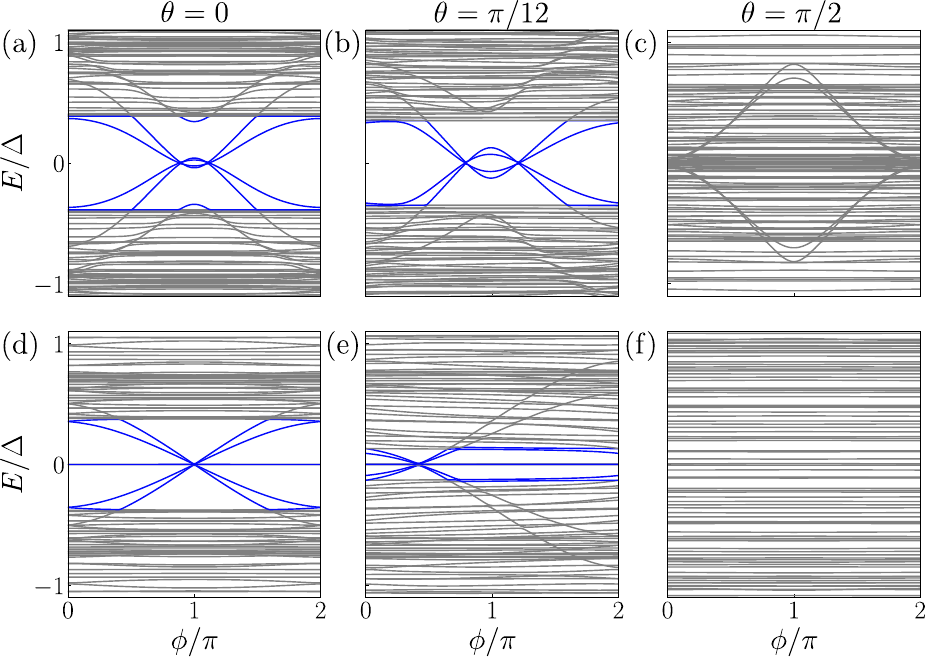}
    \caption{\textbf{Andreev spectrum of the non-interacting multichannel nanowire SNS junction with a uniform Zeeman field:}  The low-energy spectrum as a function of the phase difference $\phi$ for different orientations of the Zeeman field for: (a-c) the topologically trivial phase (B=0.5 meV) with $\mu=1$ meV and (d-f) the topologically non-trivial phase (B=1.5 meV) with $\mu=-0.5$ meV. The length of the superconductor and normal regions are taken to be $L_S=2~\mathrm{\mu m}$ and $L_N=20$ nm respectively.}
    \label{fig:SNS_spectrum_ind}
\end{figure}

When the subbands interaction is absent, each channel contributes independently without any hybridization between their corresponding spectra and the contribution of each subband is qualitatively similar to that of a pure 1D channel. In the trivial phase, this is evident by the presence of a pair of ABS coming from each subband. In the case of $\theta=0$, the spectrum is symmetric about $\phi=\pi$ as expected and additionally we observe two pairs of ABS with each pair coming from each subband. The two subbands have distinct induced gaps, therefore the quasi-continuum levels of the lowest subband coexist with the ABS of the second subband. These two pairs of ABS develop zero-crossings at the same values of $\phi$ around $\phi=\pi$ as seen in \figref{fig:SNS_spectrum_ind}(a). In the case of $\theta\ne 0$, for $\theta=\pi/12$, \figref{fig:SNS_spectrum_ind}(b) shows the usual weak asymmetry of the spectrum due to the finite $B_y$ component with a shift in positions of the zero-energy crossings around $\phi=\pi$. A more interesting behavior is seen in the case of $\theta=\pi/2$, \figref{fig:SNS_spectrum_ind}(c) shows a spectrum where the induced gap is closed completely and the energy levels are almost non-dispersive save for two dispersive levels coming from each subband. Moreover, the spectrum is completely symmetric around $\phi=\pi$ at $\theta=\pi/2$ for the independent channels. In the topological regime, the Zeeman field is taken to be $1.5$ meV such that both the subbands are in the topological phase (see \figref{fig:BdG_spectrum_ind}(a)). At $\theta=0$ (\figref{fig:SNS_spectrum_ind}(d)), due to the independent contribution of the subbands, additional pairs of ABS coexist which form a zero-crossing at $\phi=\pi$ giving rise to four pairs of MBS with two pairs coming from each subband. For $\theta=\pi/12$, the induced gap in the spectrum reduces and a strong asymmetry is created in the spectrum similar to the case of interacting channels (see \figref{fig:SNS_spectrum_int}(e)), the MBS are shifted away from $\phi=\pi$ and interestingly the MBS of each subband are formed at the same value of the junction phase-difference $\phi$ as shown in \figref{fig:SNS_spectrum_ind}(e). At $\theta=\pi/2$, \figref{fig:SNS_spectrum_ind}(f) features a low-energy spectrum where the energy levels are non-dispersive since the Fermi level no longer lies within the helical regime for the chosen values of the Zeeman field and chemical potential. The absence of spectral asymmetry for $\theta=\pi/2$ in the case of non-interacting channels indicates that no diode effect is possible without the $B_x$ component for non-interacting channels and highlights the crucial role of the inter-subband coupling in changing the properties of the JDE. 

\subsection{Current-phase relationship}
\begin{figure}
    \centering
    \includegraphics[width=\linewidth]{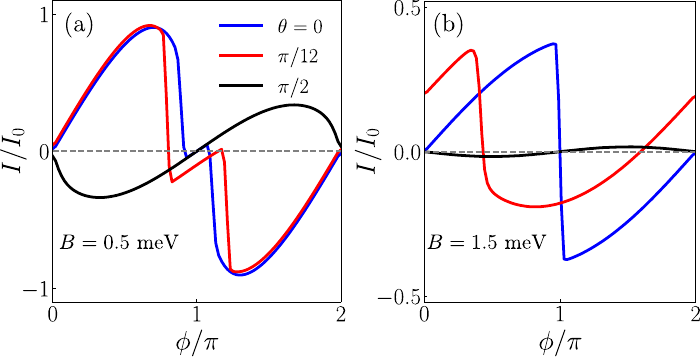}
    \caption{\textbf{Current-phase relationship of the non-interacting multichannel nanowire SNS junction with a uniform Zeeman field:}  The Josephson current at zero temperature as a function of the phase difference $\phi$ for different orientations of the Zeeman field for: (a) the topologically trivial phase (B=0.5 meV) with $\mu=1$ meV and (b) the topologically non-trivial phase (B=1.5 meV) with $\mu=-0.5$ meV. The current is normalized by $I_0=e\Delta/\hbar$.}
    \label{fig:SNS_CPR_ind}
\end{figure}
The Josephson current obtained by summing up the contribution of all discrete levels as given in \equref{current_formula} is shown in \figref{fig:SNS_CPR_ind} for the corresponding spectra in \figref{fig:SNS_spectrum_ind}. The independent contribution of the subbands results in the enhancement of the current amplitude where the critical currents are approximately double that of the interacting channels as seen from Figs. \ref{fig:SNS_CPR_int} and \ref{fig:SNS_CPR_ind}. In the trivial regime, \figref{fig:SNS_CPR_ind}(a) shows the CPR for three different values of $\theta$. The CPR for $\theta=0$ (blue curve) is symmetric about $\phi=\pi$ and exhibits a zigzag behavior which changes sign at the zero-crossings of the ABS. This demonstrates the dominant contribution of the ABS unlike the case of interacting channels where the continuum contribution lifts the sign-change of the CPR at the zero-crossings. The case of $\theta=\pi/12$ (red curve) features a CPR which is asymmetric about $\phi=\pi$ due to the non-zero $B_y$ component however the asymmetry in the critical currents $I_c^{\pm}$ is very small, thus the diode effect in the trivial regime is minimal. At $\theta=\pi/2$ (black-colored curve), the CPR is smooth and symmetric unlike the interacting-channel case which features an asymmetric CPR. In the topological regime, the CPR for $\theta=0$ features a symmetric and saw-tooth behavior with a sharp discontinuity at the zero-crossings at $\phi=\pi$ where the MBS are observed as shown in \figref{fig:SNS_CPR_ind}(b), we also note that the presence of additional ABS from the second subband results in the enhancement of the supercurrent as compared to the interacting channels. For $\theta\ne0$, the supercurrent for $\theta=\pi/12$ (red curve) in \figref{fig:SNS_CPR_ind}(b) features a  CPR with a stronger asymmetry as compared to the trivial phase, signifying again the enhancement of the diode effect by the MBS as mentioned in \secref{sec:Int_case}. We note that the asymmetry in the critical currents $I_c^+$ and $I_c^-$ is greater in the case of interacting channels for the same set of parameters, this indicates that the diode efficiency is larger for interacting channels. At $\theta=\pi/2$ the Josephson current exhibits a sinusoidal CPR with a very small critical current in spite of the non-dispersive low-energy spectrum (see \figref{fig:SNS_spectrum_ind}(f)), this indicates that the continuum levels have a small but finite contribution to the Josephson current. A striking feature of the independent-channel case is the absence of the JDE for $\theta=\pi/2$, as seen in Figs. \ref{fig:SNS_CPR_ind}(a) and (b), indicating that both the $B_x$ and $B_y$ components are required to generate the diode effect, unlike in the interacting-channel scenario.

\section{Efficiency of the Josephson diode effect}
\label{sec:efficiency}

\begin{figure}
    \centering
    \includegraphics[width=\linewidth]{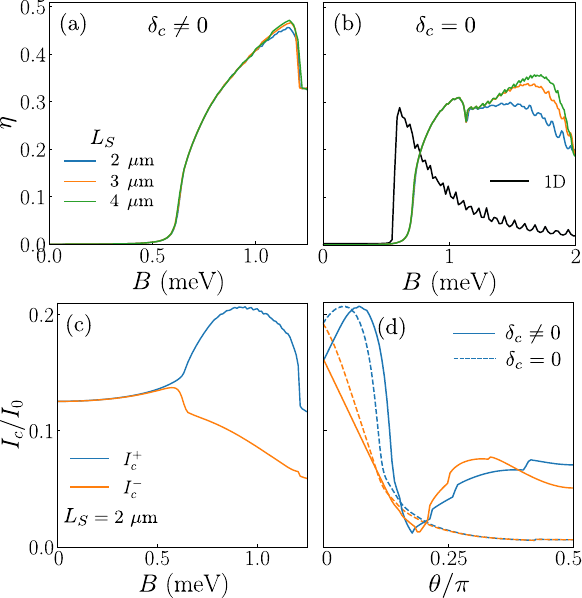}
    \caption{\textbf{JDE in the harmonically-confined nanowire Josephson junction:} Diode efficiency as a function of Zeeman field for three values of the superconducting length $L_S$ in (a) interacting channels ($\delta_c \ne 0$) and (b) independent channels ($\delta_c = 0$), with the pure 1D limit shown as a black line. Panels (c) and (d) show the asymmetry between positive and negative critical currents: (c) as a function of Zeeman energy $B$ for interacting channels, and (d) as a function of the Zeeman field orientation angle $\theta$, where the solid and dashed curves correspond to interacting and independent channels, respectively.
}
    \label{fig:efficiency_har}
\end{figure}

The asymmetry in the critical currents $I^{+}_c \ne I^{-}_c$ indicates that there is an asymmetry between the maximum Josephson current flowing in the positive and negative directions of the transport axis. This non-reciprocal behavior signifies the diode effect across the Josephson junction. The amount of non-reciprocity of the JDE is quantified by the efficiency $\eta=(I_c^+-|I_c^-|)/(I_c^+ + |I_c^-|)$ where a non-zero value of $\eta$ signifies that there is an asymmetry between the left and right-flowing Josephson current with a positive $\eta$ indicating that the right-flowing current is greater than the left-flowing current and vice versa. A larger value of $\eta$ means a JDE with a larger efficiency while a sign change indicates a reversal of the diode polarity. The efficiency and critical currents are presented in \figref{fig:efficiency_har} as a function of the Zeeman energy and the angle $\theta$ for a fully transparent junction (refer to \appref{app:tunneling_effect} for the corresponding analysis in the tunneling regime). We find that the efficiency is enhanced in the case of multichannel junctions with the enhancement being maximum in the case of interacting channels.
\figref{fig:efficiency_har}(a) shows the efficiency as a function of the Zeeman field for the interacting channels. Because of the low chemical potential ($\mu=-0.5$ meV) required to reach the helical regime of the first subband, the Fermi level lies in the vacuum below the subbands for lower values of the Zeeman field. Therefore, the lower values of the Zeeman energy do not exhibit any JDE and $\eta$ starts taking a finite value when the Zeeman energy reaches the critical field $B_c$. As $B$ increases in the topological regime, $\eta$ also increases monotonously reaching a maximum value of $\sim50\%$ and drops off significantly as $B$ lies outside the topological regime. We note that at higher Zeeman fields, the suppression of superconductivity leads to very small critical currents, which in turn results in an artificial enhancement of $\eta$. Therefore, we restrict the efficiency plot to Zeeman energies up to 1.2 meV. For the independent channels, the efficiency is lower as compared to the interacting case as seen in \figref{fig:efficiency_har}(b). For lower values of the Zeeman energy, the efficiency remains zero as mentioned above and only takes a finite non-zero value at the critical field $B_{c1}$ of the first subband. When the system is in the topological regime of the first subband the efficiency exhibits an enhancement due to the non-trivial MBS coming from the first subband, and as the Zeeman energy further increases beyond the critical field of the second subband $B_{c2}$, the efficiency is further enhanced by the presence of additional pairs of MBS coming from the second subband. Thus the maximum efficiency in this case is reached in the regime where both the subbands are in a topological phase, demonstrating the enhancement due to the presence of additional MBS. The efficiency in the case of independent channels exhibits a dependence on the superconductor length $L_S$ at larger values of the Zeeman energy due to the MBS oscillations which happen due to overlap of the MBS when $L_S<2\xi$ with $\xi$ being the localization length of the MBS. The efficiency is highest in the case of interacting channels while that of the independent channels is lower than the interacting case but still higher than the pure 1D limit. The enhancement of the JDE in the topological regime can be also seen from the asymmetry of the critical currents as shown in \figref{fig:efficiency_har}(c) where we see that the asymmetry is strongest in  the topological regime and for larger values of the Zeeman energy, the critical currents diminish due to suppression of superconductivity. The critical currents asymmetry can be also tuned by the angle $\theta$ as shown in \figref{fig:efficiency_har}(d) for both the cases of interacting and independent channels; at $\theta=0$ the asymmetry is zero since $B_y=0$ and as $\theta$ increases, $I_c^+$ becomes greater than $I_c^-$ and thus the asymmetry also increases monotonously initially due to the non-zero $B_y$ component which generates diode effect. The maximum efficiency is reached around the angle $\theta=\pi/12$, increasing the angle further results in a sharp drop of the critical currents due to the closing of the gap $\Delta_2$ at the critical angle as mentioned in \secref{sec:model}. On increasing $\theta$ further beyond the critical angle, we notice that the asymmetry becomes finite again for the interacting case and $I_c^-$ becomes greater than $I_c^+$ for $\delta_c\ne0$ which suggests that the diode polarity is now reversed. Thus the JDE polarity can be tuned by varying the angle $\theta$. A contrasting result between the interacting and independent channels is seen at $\theta=\pi/2$; the independent-channel case exhibits no critical currents asymmetry for larger values of $\theta$ while a small but finite asymmetry persists in the interacting-channel case which further demonstrates that the JDE can be realized through the transverse field alone in the interacting-multichannel nanowire.

\section{Finite temperature effects}
\label{sec:finite_temp}

\begin{figure}
    \centering
    \includegraphics[width=\linewidth]{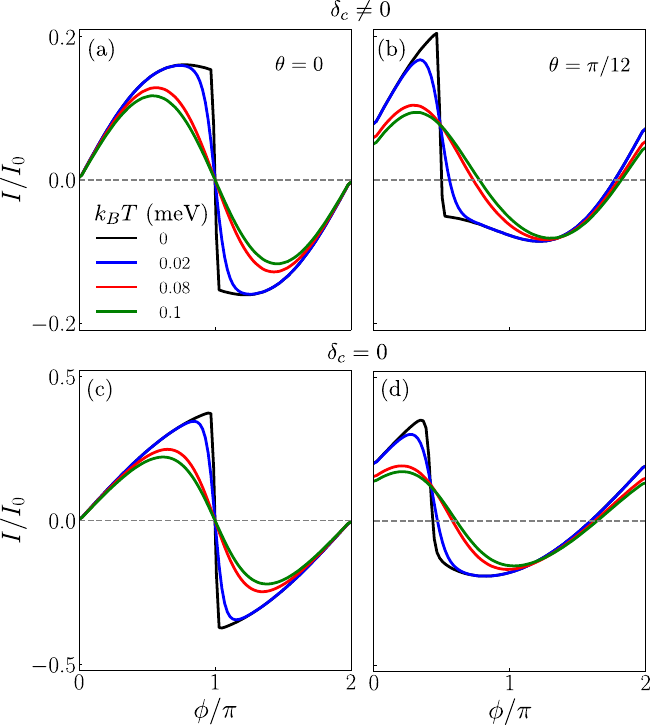}
    \caption{\textbf{Variation of the CPR for different values of the temperature:}
    The Josephson current in the topological phase shown as a function of the phase-difference at different temperatures for: (a,b) the interacting-channel case and (c,d) non-interacting channel case. The figures in the two columns correspond to $\theta=0$ and $\theta=\pi/12$ respectively.  Here, $L_S=2~\mathrm{\mu m}$, $L_N=20$ nm, and $\mu=-0.5$ meV.}
    \label{fig:temp_effect}
\end{figure}
The results that we have shown so far are calculated at zero temperature, however, a finite temperature can have a significant impact in modifying the properties of the Josephson current and the diode efficiency. To investigate the finite temperature effects we compute the Josephson current using \equref{current_formula} and choose a finite non-zero $k_BT$ value. In \figref{fig:temp_effect} we present the CPRs for different values of $k_BT$ for both the interacting and independent channels in the topological regime. \figref{fig:temp_effect}(a) and (b) show the CPR for the interacting-channel case for $\theta=0$ and $\theta=\pi/12$ respectively. The effect of finite temperature on the supercurrent is immediately visible by the suppression of the critical currents with increasing temperature due to the diminishing of the $s-$wave superconducting gap. \figref{fig:temp_effect}(a) shows the comparison of the CPRs at $\theta=0$ for different temperatures. We notice that the shape of the CPR gets modified and the saw-tooth behavior of the zero-temperature supercurrent gets wiped out even under a small finite temperature due to thermal smearing effect. As the temperature is increased further, the critical current decreases and becomes sinusoidal at higher temperatures. The effects of finite temperature are even more significant in the case of $\theta=\pi/12$, in addition to the smearing of the sharp discontinuity of the zero-temperature supercurrent, the asymmetry of the critical currents $I_c^{\pm}$ is also reduced with increasing temperature as seen in \figref{fig:temp_effect}(b). This implies that the efficiency of the JDE is suppressed by finite temperature effects. The results for the finite-temperature effects in the independent-channel case are similar to that of the interacting case, we notice that the symmetric CPR in \figref{fig:temp_effect}(c) exhibits thermal smearing and reduced critical currents with temperature and the asymmetric CPR for $\theta=\pi/12$ in \figref{fig:temp_effect}(d) features a reduced asymmetry of the critical currents as observed in the interacting-channel case. This indicates that these effects are generic and arise purely due to finite temperature. 

\begin{figure}
    \centering
    \includegraphics[width=\linewidth]{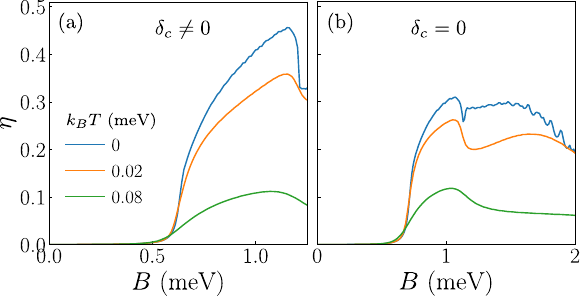}
    \caption{\textbf{Variation of the diode efficiency with temperature:}
    Comparison of the JDE efficiency at zero and finite temperatures for the (a) interacting-channel case and (b) independent-channel case. Here, $\theta=\pi/12$ and $L_S=2~\mathrm{\mu m}$.}
    \label{fig:eff_temp}
\end{figure}
The reduction of critical currents asymmetry results in the suppression of the JDE with increasing temperature. Indeed, this is observed in \figref{fig:eff_temp} where we compare the efficiency as a function of the Zeeman energy for different temperatures for both the interacting and independent channels. The efficiency in the interacting-channel case (\figref{fig:eff_temp}(a)) has a maximum value of $45\%$ in the zero-temperature limit for the chosen superconductor length of $L_S=2~\mathrm{\mu m}$ and for the finite temperature $k_BT=0.02$ meV the maximum efficiency reduces to $38\%$ and keeps reducing as the temperature increases. Similarly, for the independent-channel case (\figref{fig:eff_temp}(b)) the efficiency exhibits a similar variation with temperature where the maximum efficiency is $30\%$ at zero temperature and gets reduced with increasing temperature.

\section{Effect of occupying multiple spinful subbands}
\label{sec:3_subbands}
\begin{figure}
    \centering
    \includegraphics[width=\linewidth]{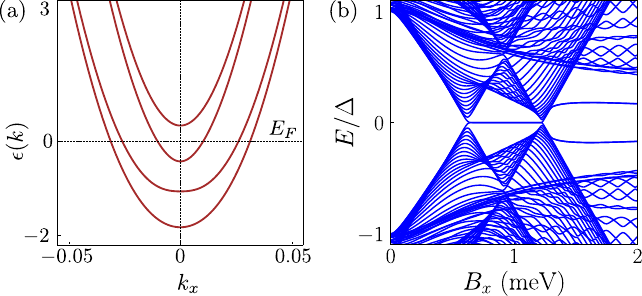}
    \caption{\textbf{Low-energy spectrum with three occupied spinful bands in the multichannel Rashba superconductor nanowire:}
    (a) The normal-state dispersion showing the Fermi level cutting through three spin-split bands for the parameters: $B=0.7$ meV and $\mu=1$ meV. (b) The corresponding low-energy spectrum of the multichannel Rashba superconductor nanowire as a function of the Zeeman energy. and the SNS junction low-energy spectrum as a function of the phase difference $\phi$ for (c) $\theta=0$ and (d) $\theta=\pi/12$, respectively.}
    \label{fig:3subb_1}
\end{figure}

\begin{figure}
    \centering
    \includegraphics[width=\linewidth]{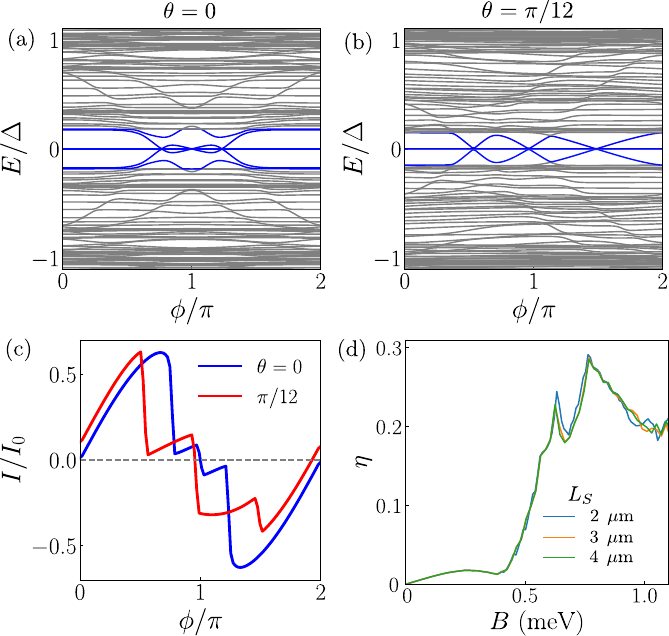}
    \caption{\textbf{Properties of the SNS junction with three occupied spinful bands:} Low-energy BdG spectrum of the SNS junction as a function of the phase difference $\phi$ for (a) $\theta=0$ and (b) $\theta=\pi/12$, respectively. (c) The CPR corresponding to (a) and (b), and (d) the efficiency of the JDE for $\theta=\pi/12$ as a function of the Zeeman field for different values of $L_S$.}
    \label{fig:3subb_2}
\end{figure}

The chemical potential can be tuned such that an odd number of spinful subbands is occupied. In \figref{fig:3subb_1}(a) the chemical potential is taken to be $\mu=1$ meV such that three helical subbands are occupied at lower values of the Zeeman field. The low-energy spectrum of the interacting-channel case in this regime as a function of the Zeeman field is qualitatively similar to the case of one occupied-subband (see \figref{fig: dispersion_topology}(c)) as shown in \figref{fig:3subb_1}(b) and exhibits the spectral even-odd effect with MBS emerging beyond the critical field which later hybridize to two trivial non-zero energy states as the Zeeman field increases since the occupancy changes from three to four subbands. The low-energy spectrum of the SNS junction as a function of the phase difference is, however, starkly different to the one-spinful-band occupancy case. For the symmetric spectrum ($\theta=0$), in addition to the non-dispersive MBS at the nanowire ends and the MBS at $\phi=\pi$  the two lowest ABS develop two additional zero-energy crossings on either side of $\phi=\pi$ which can be understood as the trivial crossings coming from the first two bands \cite{Stanescu_2011}; \figref{fig:3subb_2}(a). This is seen more clearly in \figref{fig:3subb_2}(b) which shows the asymmetric spectrum for $\theta=\pi/12$ where the MBS are shifted away from $\phi=\pi$ due to the $B_y$ component and the two trivial zero-energy crossings come from the other two ABS branches. The CPR corresponding to \figref{fig:3subb_2}(a) and (b) is shown in \figref{fig:3subb_2}(c) and features zig-zag patterns due to the trivial crossings of the first two subbands and a kink-like behavior at the crossing corresponding to the MBS. We also notice that for the interacting channels, the critical currents asymmetry is reduced when multiple subbbands are occupied. This indicates that the overall efficiency is less as compared to the case when one spinful subband is occupied. Indeed we see in \figref{fig:3subb_2}(d) that the maximum efficiency in this case is $\sim 30\%$ in the topological regime while the efficiency in the trivial regime is negligible.

\section{Robustness of Diode effect with different confinement potentials}
\label{sec:rect_case}

\begin{figure}
    \centering
    \includegraphics[width=\linewidth]{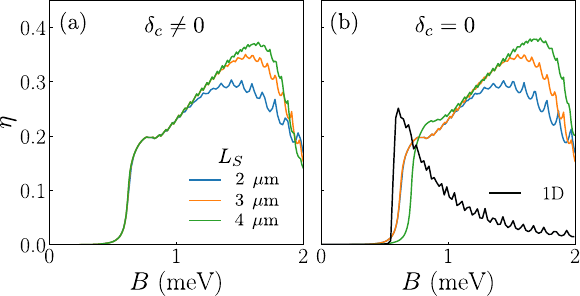}
    \caption{\textbf{Diode efficiency of the rectangular nanowire Josephson junction:}
    The JDE efficiency shown as a function of the Zeeman field for three values of the superconductor length $L_S$ in the case of: (a) interacting channels ($\delta_c\ne0$) and (b) independent channels ($\delta_c=0$) with the pure 1D limit (black curve).}
    \label{fig:efficiency_rect}
\end{figure}

In this section, we show that the characteristics of the Rashba nanowire Josephson junctions are qualitatively same in nanowires with other types of confinement potentials and geometry. We compare the results from the harmonically-confined nanowire to a rectangular Rashba nanowire of dimensions $L_x,L_y$ and $L_z$ with hardwall confinement along the $z$-direction enforced by restricting the nanowire lengths as $L_z\ll L_y\ll L_x$. This results in subbands sitting at discrete energy levels $E_n=n^2\pi^2\hbar^2/(2m^*L_y)$ with $n=1,2\ldots$, and the corresponding inter-subband coupling is now given by $\delta_c=8\hbar\alpha/(3L_y)$. We find that the low-energy spectrum and topological properties in this case share a qualitative similarity to the harmonically-confined nanowire. Moreover, the effects of the Zeeman field and its relative orientation on the spectrum and CPR are consistent with the results observed earlier, further demonstrating that these results are robust for any generic multichannel system. We present here only the efficiency of the JDE for the interacting and non-interacting cases in \figref{fig:efficiency_rect} as a function of the Zeeman field. \figref{fig:efficiency_rect}(a) and (b) show the efficiency for $\theta=\pi/12$ as a function of the Zeeman energy for the case of interacting and independent channels respectively. The efficiency depends on the superconductor length and exhibits oscillations with the increasing Zeeman field even at $L_S=3~\mathrm{\mu m}$ due to the overlap of the non-local MBS, this suggests that the Majorana localization length is affected by the confinement. As the length $L_S$ increases gradually, the oscillation behavior decreases due to the reduced non-locality of the MBS which also increases the efficiency, reaching a maximum value of $\sim40\%$ for $L_S=4~\mathrm{\mu m}$. The maximum efficiency in the case interacting and independent channels is equal in a rectangular Rashba nanowire JJ and is slightly lower as compared to the harmonically-confined Rashba nanowire for the same set of parameters, however the diode efficiency is still enhanced for the multichannel as compared to the pure 1D channel junction.

\section{Conclusions}
\label{sec:conclusion}
In this work, we have theoretically characterized the JDE in multichannel semiconductor-superconductor hybrid nanowire Josephson junctions, focusing on the interplay between a tilted Zeeman field and transverse confinement. Our results show that inter-subband coupling, mediated by spin-orbit interactions, qualitatively reshapes the topological phase diagram: the topological phase is realized only when an odd number of subbands are occupied, thereby confining the Majorana-supporting regime to a finite window of Zeeman energy. We further demonstrate that subband hybridization induces a pronounced asymmetry in the ABS spectrum with respect to the superconducting phase difference, which underlies the emergence of nonreciprocal supercurrents. A key consequence of this multichannel physics is the appearance of a finite JDE even when the Zeeman field is aligned purely along the spin-orbit direction-- a feature absent in strictly one-dimensional or independent-channel limits. By comparing single-channel, independent-channel, and interacting-channel regimes, we find that the diode efficiency is significantly enhanced in multichannel junctions. In the interacting-channel case, this enhancement originates from increased spectral asymmetry due to inter-subband coupling, whereas in the independent-channel limit it arises from the presence of additional Majorana bound states in the topological regime. The qualitative consistency of these features across different confinement potentials further underscores the robustness of the underlying mechanism.

More broadly, our results highlight that going beyond idealized single-channel descriptions is essential for understanding physics of realistic hybrid nanowire devices. The multichannel nature of these systems not only modifies quantitative aspects of transport but also enables qualitatively distinct behavior driven by subband hybridization. In this context, controlling the subband structure and inter-subband coupling provides a natural route for tuning nonreciprocal superconducting responses. Our findings offer a framework for interpreting experiments in multichannel nanowires and point toward the potential of leveraging multiband effects to design superconducting devices with enhanced nonreciprocal functionalities. We expect that the interplay between multichannel topology and nonreciprocity explored here will stimulate further studies and contribute to the development of superconducting circuits with enhanced directional functionalities.
\\

\noindent\emph{Note added.} During the preparation of this manuscript, a related work (arXiv:2510.05772) appeared. We find qualitative agreement with their results, particularly regarding the enhancement of the diode efficiency, while the underlying model and calculation setup remain distinct.

\subsection*{Acknowledgments}
SKG acknowledges financial support from Anusandhan National Research Foundation (ANRF) erstwhile Science and Engineering Research Board (SERB), Government of India via the Startup Research Grant: SRG/2023/000934 and from IIT Kanpur via the Initiation Grant (IITK/PHY/2022116). AS and SKG utilized the \textit{Andromeda} server at IIT Kanpur for numerical calculations. We acknowledge Sagar Santra and Dibyendu Samanta for many enlightening discussions.


\appendix

\section{Josephson diode effect in long multichannel junctions}
\label{app:long_junction}
\begin{figure}
    \centering
    \includegraphics[width= \linewidth]{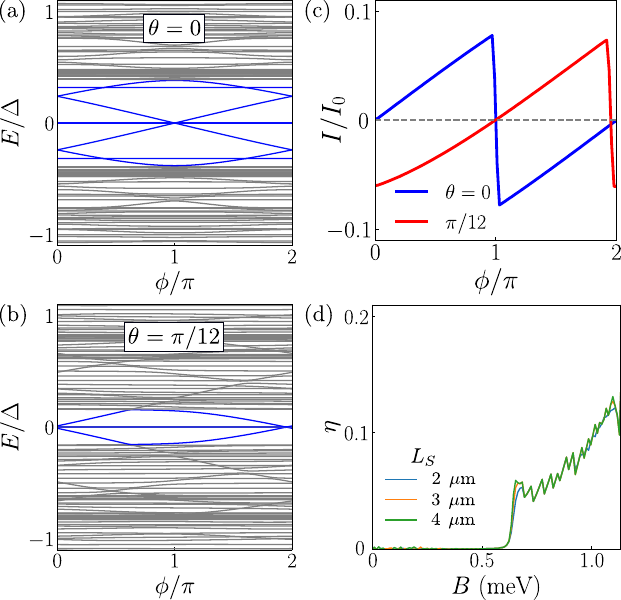}
    \caption{\textbf{JDE in the long-junction limit:} The low-energy spectrum of the interacting multichannel Josephson junction in the topological phase ($B=1$ meV) with the length of the normal region taken as $L_N=500~\mathrm{n m}$ for: (a) $\theta=0$ and (b) $\theta=\pi/12$. (c) The CPR corresponding to the spectrum in (a) and (b). (d) The JDE efficiency as a function of the Zeeman energy for different values of $L_S$ for $\theta=\pi/12$. In panels (a-c) the superconductor length is taken as $L_S=2~\mathrm{\mu m}$ and the chemical potential is fixed at $\mu=-0.5$ meV in all panels.}
    \label{fig:long_junction}
\end{figure}
We have limited our discussion to the case of short junctions in the main text by considering normal regions of length $L_N=20$ nm. In this section we provide additional information regarding the behavior of the JDE in when junctions with longer N regions are taken into account. We present the results for normal region length $L_N=500$ nm in \figref{fig:long_junction}, the low-energy spectrum in Figs. \ref{fig:long_junction}(a) and (b) features a spectrum that is greatly modified from the short-junction case, the spectrum is mostly non-dispersive in the quasi-continuum and continuum regime with additional energy levels introduced in both the subgap and quasi-continuum regime as more energy states are hosted by the longer normal region. The symmetric spectrum for $\theta=0$ shows that the shape of four lowest levels that that form MBS at $\phi=\pi$ have modified to straight lines as opposed to curved levels in the short-junction limit. This change in the spectral feature has direct effect on the shape of the corresponding CPR. In addition to MBS levels, the junction also hosts additional dispersive and non-dispersive trivial ABS levels. The finite non-zero angle $\theta=\pi/12$ creates an asymmetry in the low-energy spectrum, shifting the MBS far away from $\phi=\pi$ close to $2\pi$. The additional trivial subgap levels also get pushed towards the quasi-continuum regime leaving only the MBS levels within the induced gap.

The modified features of the energy spectrum have a significant impact on the CPR of the long junction as shown in Fig. \ref{fig:long_junction}(c). The CPR for $\theta=0$ features a saw-tooth shape with a much reduced critical current. This is expected since in the long junction limit the continuum levels are non-dispersive and the contribution to the CPR comes mainly from the MBS levels. The non-reciprocal CPR for $\theta=\pi/12$ shows that the current changes sign discontinuously at the shifted-MBS crossing near $\phi=2\pi$ and the asymmetry of the critical currents is smaller than the short-junction limit. This indicates a JDE with a reduced efficiency as seen in \figref{fig:long_junction}(d), we first observe that the efficiency is zero for low values of the Zeeman field as the Fermi level lies in the vacuum owing to the low chemical potential and as the Zeeman field increases the efficiency becomes finite but small in the topological regime. The long junction has a maximum efficiency of around $30\%$ for a small interval of $B$ and beyond this peak the efficiency exhibits spurious enhancements due to the very small critical currents. Thus we have showed that the enhancement of the JDE is not robust in the long-junction limit of the multichannel nanowires.

\section{Dispersion asymmetry driven solely by the transverse field}
\label{app:transverse_field_asymmetry}

\begin{figure} 
    \centering
    \includegraphics[width= 0.9\linewidth]{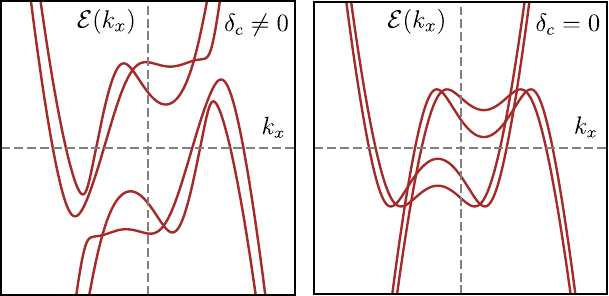}
    \caption{\textbf{Bogoliubov quasiparticle dispersion of the proximitised multichannel Rashba nanowire when the Zeeman field is purely along the SOC direction ($\theta=\pi/2$):} (a) The asymmetric multiband dispersion in the case of interacting channels ($\delta_c\ne0$) and (b) Symmetric dispersion in the case of independent-channels ($\delta_c=0$).}
    \label{fig:asymmetry_pi_2}
\end{figure}

In the main text we have emphasized the fact that the JDE can be realized through the transverse field alone in the case of interacting-channel nanowire, here we show that this is a unique property where the asymmetry arises from the interplay of the inter-subband coupling with the $B_y$ component of the Zeeman field. In \figref{fig:asymmetry_pi_2}, we present the energy-momentum dispersion of the multichannel Rashba nanowire in the case of interacting ($\delta_c\ne0$) and independent ($\delta_c=0$) channels when the external Zeeman field lies wholly along the SOC-axis by setting $\theta=\pi/2$. For the interacting channels, the spectrum exhibits an asymmetry about $k_x=0$ leading to the formation of finite-momentum Cooper pairs which is crucial for observing diode effect. In the case of the independent channels, the spectrum is symmetric about $k_x$ with the electron-like and hole-like bands possessing different points of symmetry along $k_x$. Thus the asymmetric Fulde-Ferrell-Larkin-Ovchinnikov (FFLO) state is not realized in this case and hence no diode effect is possible.

\section{Josephson diode effect in the tunneling regime }
\label{app:tunneling_effect}
\begin{figure}
    \centering
    \includegraphics[width=\linewidth]{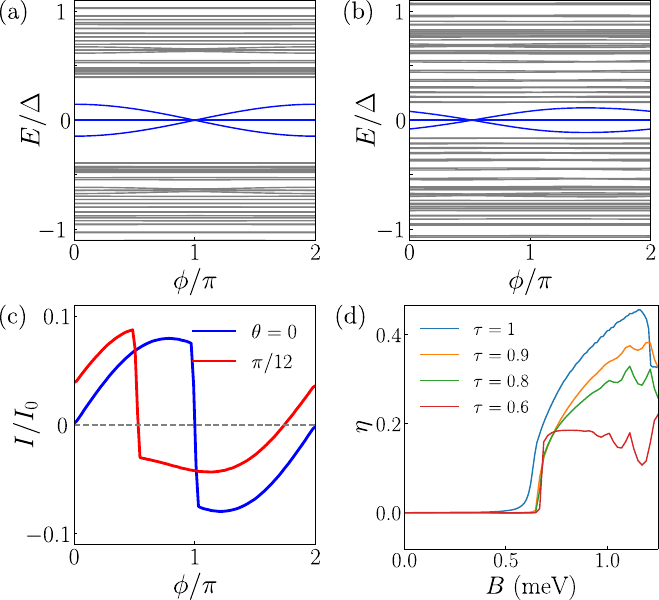}
    \caption{\textbf{Features of the JDE for the interacting-channel case in the tunneling regime:} The low-energy spectrum in the tunneling regime ($\tau=0.8$) for (a) $\theta=0$ and (b) $\theta=\pi/12$. (c) The current-phase relationship corresponding to (a) and (b), and (d) the efficiency of the JDE at $\theta=\pi/12$ shown as a function of the Zeeman energy for several values of the transmission coefficient $\tau$. Here, $\mu=-0.5$ meV, $B=1$ meV, $L_S=2~\mathrm{\mu m}$ and $L_N=20$ nm.
    }
    \label{fig:tunneling_regime}
\end{figure}
The diode effect is investigated in the main text in the limit of a fully transparent ($\tau=1$) junction. Here, we extend this investigation by providing additional results for the tunneling regime, where a potential barrier at the junction reduces the transmission parameter to a value $\tau<1$. The junction transparency is incorporated by multiplying the transmission coefficient with the hopping term $\tau T_{ij}$ in \equref{eq:SNS_hamil} at the nearest-neighbor sites between the superconductor and normal regions. Recent studies in the pure 1D JJs reveal that the critical currents crucially depend on the transmission coefficient; in the tunneling regime the critical current exhibits a re-entrant behavior where it vanishes below the critical field and re-appears in the topological regime \cite{Cayao_2017}. This gives rise to a JDE which is dominated by the MBS in the tunneling regime. In the interacting multichannel junction, the symmetric low-energy spectrum for $\theta=0$ when $\tau=0.8$ shows that the four lowest states that form MBS have become less dispersive with $\phi$ and the continuum states are almost non-dispersive, showing that the contribution to the Josephson current comes only from the MBS. The spectrum for $\theta=\pi/12$ also show similar behaviors along with the $B_y$-induced asymmetry; \figref{fig:tunneling_regime}(a) and (b). Since the continuum is not contributing significantly to the the supercurrent, the corresponding supercurrent features a saw-tooth CPR with a reduced critical current for $\theta=0$ and a non-reciprocal CPR for $\theta=\pi/12$ with a reduced asymmetry between the critical currents $I_c^{\pm}$, indicating a reduction in the JDE efficiency in the tunneling regime. Indeed, the efficiency exhibits a strong dependence on the transmission coefficient as seen in \figref{fig:tunneling_regime}(d) where the efficiency is plotted as a function of the Zeeman energy. The efficiency is maximum in the case of a fully transparent ($\tau=1$) junction and decreases with $\tau$ and reaches a saturated value at $\tau=0.5$. We also notice that the efficiency exhibits an oscillation with the Zeeman energy in the tunneling regime, this is similar to the case of a pure 1D JJ, where the period of MBS oscillation is doubled in the tunneling regime \cite{Cayao_2017}. In the case of multichannel junctions, the oscillation is absent in the transparent limit but appears in the tunneling regime.

\bibliography{ref}

\end{document}